# Detuning related coupler kick variation of a superconducting nine-cell 1.3 GHz cavity

Thorsten Hellert[*] and Martin Dohlus
*DESY, Notkestrasse 85, 22603 Hamburg, Germany*

Superconducting TESLA-type cavities are widely used to accelerate electrons in long bunch trains, such as in high repetition rate free electron lasers. The TESLA cavity is equipped with two higher order mode couplers and a fundamental power coupler (FPC), which break the axial symmetry of the cavity. The passing electrons therefore experience axially asymmetrical coupler kicks, which depend on the transverse beam position at the couplers and the rf phase. The resulting emittance dilution has been studied in detail in the literature. However, the kick induced by the FPC depends explicitly on the ratio of the forward to the backward traveling waves at the coupler, which has received little attention. The intention of this paper is to present the concept of discrete coupler kicks with a novel approach of separating the field disturbances related to the standing wave and a reflection dependent part. Particular attention is directed to the role of the penetration depth of the FPC antenna, which determines the loaded quality factor of the cavity. The developed beam transport model is compared to dedicated experiments at FLASH and European XFEL. Both the observed transverse coupling and detuning related coupler kick variations are in good agreement with the model. Finally, the expected trajectory variations due to coupler kick variations at European XFEL are investigated and results of numerical studies are presented.

DOI: 10.1103/PhysRevAccelBeams.21.042001

## I. INTRODUCTION

Single pass free electron lasers (FEL) are the state-of-the-art particle accelerators to generate high brilliance photon pulses [1]. At the FEL user facilities FLASH (Free-Electron Laser in Hamburg) [2,3] and European XFEL (European X-Ray Free-Electron Laser) [4–6], the driving electron bunches are accelerated in superconducting radio-frequency (rf) resonators based on the TESLA (TeV-Energy Superconducting Linear Accelerator) [7] technology. The advantage of superconducting rf cavities is the ability to provide high bunch repetition rates.

The TESLA cavity is a 9-cell standing wave structure of about 1 m length whose fundamental transverse magnetic mode resonates at 1.3 GHz. It is equipped with a higher order mode (HOM) coupler at each end of the cavity [8] in order to extract undesired field excitations. The fundamental power coupler (FPC) is installed horizontally at the downstream end of the cavity and connects the cavity to its power source.

Couplers break the axial symmetry of the cavity [9]. Their impact on the beam is illustrated in Fig. 1, where tracking results are shown for a particle which enters the cavity on axis with an initial beam energy of 120 MeV. Plotted are the longitudinal (top), vertical (mid), and horizontal (bottom) momentum as a function of the longitudinal coordinate $z$. The significant change of transverse momenta at the coupler regions are referred to as coupler kicks. The sinusoidal variation inside the cavity is related to axially symmetrical rf focusing [10], since the upstream HOM coupler kicks the beam off axis.

Coupler kicks depend on the rf phase and therefore distort different longitudinal slices of the beam by a different amount, which results in an increase of the projected emittance [11]. A large volume of published studies describe the effect of coupler kicks on emittance dilution [12–14] and its mitigation for different geometries of the superstructure [15–19]. However, the existing literature on coupler kicks largely ignores the role of different field configurations related to the forward and backward traveling waves in the FPC. This work aims to close that gap.

Tracking in Fig. 1 was done using different cavity field configurations, while the amplitude and phase of the accelerating field was kept constant. The blue color corresponds to a purely backward wave, thus a wave traveling from the cavity into the waveguide. The blue color corresponds to the opposite, where the electromagnetic field is solely defined by the forward traveling wave from the waveguide into the cavity. The red color corresponds to a standing wave, where both traveling waves have the same magnitude.

[*]thorsten.hellert@desy.de

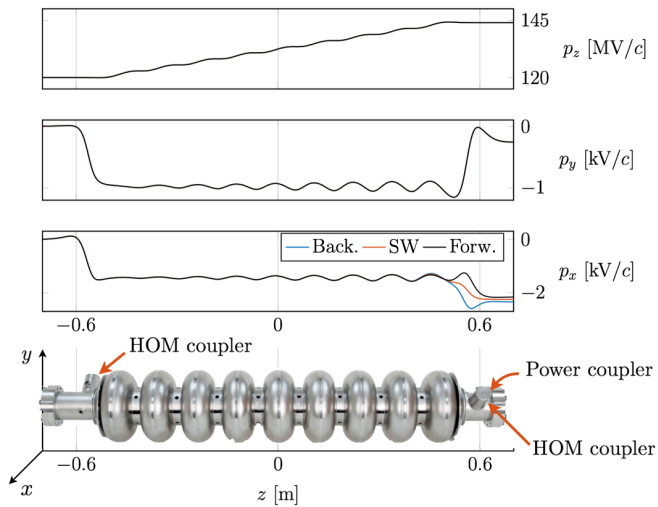

FIG. 1. Tracking results for a particle which enters a TESLA cavity on axis with an initial beam energy of 120 MeV. Plotted are the longitudinal (top), vertical (mid) and horizontal (bottom) momentum as a function of the longitudinal coordinate $z$. The accelerating gradient is 24 MV/m. The significant change of transverse momenta at the coupler positions are related to coupler kicks. The sinusoidal variation inside the cavity is related to axially symmetrical rf focussing. Tracking is calculated for a purely forward traveling wave (black), a standing wave (red), and a backward wave (blue). Only the downstream horizontal coupler kick depends on the mode of cavity operation.

Figure 1 shows that the downstream coupler kick in the horizontal plane depends considerably on the reflection factor of the two traveling waves, whereas the energy gain and the vertical momentum are hardly affected. This is due to the fact that the FPC is mounted horizontally on the cavity and that the fields far away from the coupler are independent of the direction of the traveling waves.

The quality factor of the TESLA cavity is on the order of $10^{10}$ and wall losses can be neglected. In the absence of beam loading, the complex reflection factor is therefore determined by the detuning of the cavity. Due to the high loaded quality factor of $> 10^6$, a small deformation of the cavity shape, for example because of dynamic Lorentz forces [20] or microphonics [21], results in considerable detuning of the cavity. Variations of the detuning consequentially produce variations of the coupler kicks.

The present study aims to explore the relationship between cavity detuning and coupler kicks. It provides a novel approach to quantify detuning related coupler kick variations by separating the impact of the standing and the traveling wave. Particular attention is furthermore given to the role of the penetration depth of the coaxial FPC antenna, thus the value of the loaded quality factor of the cavity.

This paper is organized in the following way: it begins with an introduction to the TESLA cavity and its couplers in Sec. II. Section III provides the mathematical principles to calculate general electromagnetic field configurations from a given field map. The main idea of discrete coupler kicks is developed in Sec. IV and quantified for different penetration depths of the FPC antenna in Sec. V. In Sec. VI we present the implementation of a linear beam dynamics model relying on discrete coupler kicks and proof of its applicability by comparison with particle tracking. Comparisons with dedicated experiments at FLASH and European XFEL are given in Sec. VII. Finally, the developed coupler kick model is applied in Sec. VIII for beam dynamics simulations of European XFEL.

## II. TESLA CAVITY

The TESLA cavity [7] is a 9-cell standing wave structure of about 1 m length whose lowest TM mode resonates at $f_0 = 1.3$ GHz. The cavity is built from solid niobium and is cooled by superfluid helium at 2 K. The fundamental advantage of superconducting cavities as compared to normal conducting cavities is the low surface resistance of about 10 nΩ at 2 K. This allows for high wall currents with little heat losses. The quality factor $Q_0$ of a cavity is defined as the ratio of the energy stored in the cavity $U$ to the energy dissipated in the cavity walls $P_{\text{diss}}$ per rf cycle,

$$Q_0 = \frac{2\pi f_0 U}{P_{\text{diss}}}, \qquad (1)$$

which corresponds to the ratio of the resonance frequency to the width of the resonance curve. The typical quality factors of normal conducting cavities are in the range of $10^4$–$10^5$ while for TESLA cavities $Q_0 > 10^{10}$. A schematic drawing of a TESLA cavity is shown in Fig. 2.

The bunched electron beam excites eigenmodes of a variety of frequencies in the cavity. If the frequency of the mode exceeds the resonance frequency of the fundamental mode, it is called higher order mode (HOM). Due to the high loaded quality factor, the damping time constants of these modes are large compared to the typical bunch spacing. In order to prevent multibunch instabilities and beam breakup [22,23], additional measures to damp HOMs have to be taken. For this reason a HOM-coupler [8] is mounted at each end of the cavity in order to extract undesired HOM field energy. A 1.3 GHz notch filter is incorporated to prevent energy extraction from the fundamental mode. Upstream and downstream HOM coupler are oriented at 115° with respect to each other, as illustrated in Fig. 3. The geometry of the HOM coupler is the same at FLASH and European XFEL.

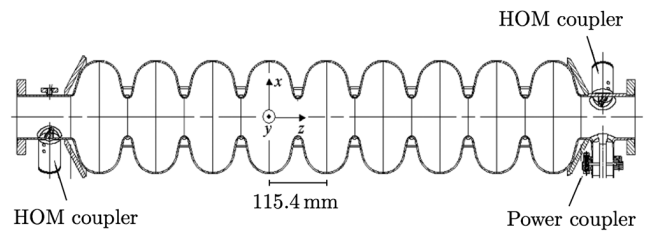

FIG. 2. Longitudinal cross section of a TESLA cavity. The beam direction is from left to right, thus the fundamental power coupler is located at the downstream end of the cavity.

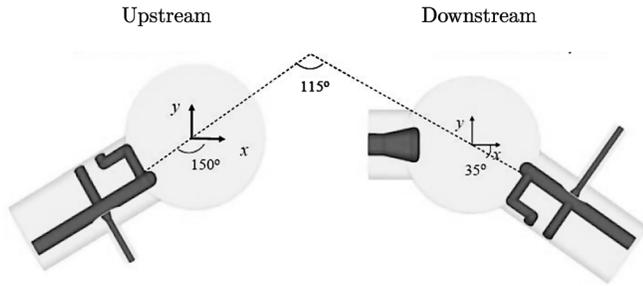

FIG. 3. Geometry and orientation of the higher order mode (HOM, upstream, and downstream) and fundamental power coupler (FPC, downstream).

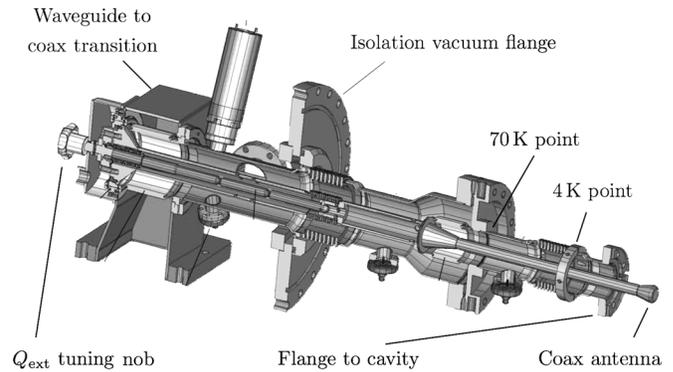

FIG. 4. Schematic drawing of the TTF-3 fundamental power coupler of the TESLA cavity. The waveguide (lower left end) connects the cavity with the rf power source and is at room temperature. A remote controlled stepper motor allows one to move the position of the coaxial antenna in the cavity beam pipe (right end).

The fundamental power coupler (FPC) transports the high power rf from a warm, air-filled waveguide system through a coaxial connection into the cold cavity. The amount of power coupled from the waveguide into the cavity and vice versa is characterized by the external quality factor $Q_{ext}$ of the cavity. The loaded quality factor $Q_L$ characterizes both the wall losses and the external power losses through the coupler and it follows from Eq. (1) that

$$\frac{1}{Q_L} = \frac{1}{Q_{ext}} + \frac{1}{Q_0}, \qquad (2)$$

where for superconducting cavities $Q_{ext} \ll Q_0$. The loaded quality factor is therefore essentially determined by the external losses through the coupler. Different coupler specifications were developed [24] in order to account for different operational conditions.

Figure 4 shows a simplified schematic of the TTF-3 power coupler used at FLASH and European XFEL. It is based on the TESLA Linear Collider [25] design and optimized for pulsed rf operation with high accelerating gradients and a duty cycle of about 1%. The rf input power is about 200 kW for a beam current of 5 mA. By moving the inner conductor of the coaxial line in a range of 10 mm, the loaded quality factor $Q_L$ can be varied in the nominal range $10^6$–$10^7$ to allow not only for different beam loading conditions [26], but also to facilitate an in-situ high power processing of the FPCs [7]. The cavity bandwidth at the typical value of $Q_L = 3 \times 10^6$ at FLASH is about 430 Hz. European XFEL operates at $Q_L = 4.6 \times 10^6$ with a resulting bandwidth of about 280 Hz.

At FLASH and European XFEL, a 1.3 GHz TESLA cavity based injector module accelerates the beam off crest in order to impose an energy chirp along the bunch for the needs of further longitudinal bunch compression.

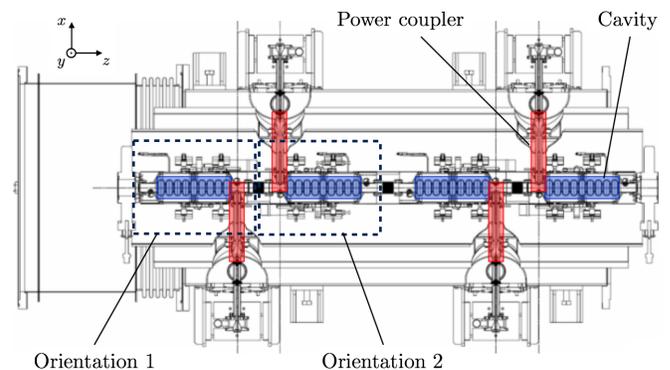

FIG. 5. Schematic drawing of the third harmonic module at FLASH. Four cavities with alternate coupler orientation with respect to the beamline are installed in one cryomodule.

A following third harmonic system, operating at 3.9 GHz, decelerates the beam and thereby linearizes the longitudinal phase space of the bunches.

The 3.9 GHz cavity [27,28] design is similar to a scaled version of the 1.3 GHz TESLA cavity. An alternate coupler orientation for each second cavity is used to compensate the coupler kick partially. At FLASH each second cavity is rotated by 180° around the y-axis [29], as shown in Fig. 5. The European XFEL uses a rotation of 180° around the z-axis [30], which leads to a cancellation of the offset independent part of the coupler kick.

The power coupler antenna penetration depth in the 3.9 GHz cavity is fixed. In order to change $Q_L$, a 3-stub tuner is installed in the waveguides. The cavity geometry in the proximity of the beam does therefore not vary for different $Q_L$.

## III. FIELD CALCULATIONS

In order to investigate coupler kicks numerically, precise knowledge of the electromagnetic fields is required. This section provides the mathematical principles which are needed to generalize the field configuration from a given traveling wave field map.

The time harmonic field at any point $\vec{r}$ in a perfect electric conducting cavity that is equipped with couplers can be written as:

$$\vec{E}(\vec{r},t) = \Re\{\vec{\mathbf{E}}(\vec{r}) \cdot e^{i\omega t}\}$$
$$\vec{B}(\vec{r},t) = \Re\{\vec{\mathbf{B}}(\vec{r}) \cdot e^{i\omega t}\}$$
$$\vec{\mathbf{E}}(\vec{r}) = A_f e^{i\phi_f} \vec{\mathbf{E}}_0(\vec{r}) + A_b e^{i\phi_b} \vec{\mathbf{E}}_0^*(\vec{r})$$
$$\vec{\mathbf{B}}(\vec{r}) = A_f e^{i\phi_f} \vec{\mathbf{B}}_0(\vec{r}) - A_b e^{i\phi_b} \vec{\mathbf{B}}_0^*(\vec{r}). \quad (3)$$

$A_{[f/b]}$ and $\phi_{[f/b]}$ are the amplitude and phase of the forward and backward traveling wave to/from the power coupler, respectively, $\omega$ the angular frequency of excitation and $\vec{\mathbf{E}}_0$, $\vec{\mathbf{B}}_0$ the forward solution for the electric and magnetic field component of the excited mode. The bold letters indicate complex values.

A standing wave field is stimulated if $|A_b| = |A_f|$. Ignoring wall losses this condition is fulfilled in absence of a beam current and can be easily realized for EM field calculation either with port stimulation or by an eigenmode solver [15].

$|A_b| \neq |A_f|$ corresponds to a net-energy-flow from/to the cavity. Ignoring wall losses, this is the case if the stored energy in the cavity is changed, for example while filling the cavity, or through beam loading. In order to account for an energy flow, a traveling wave is required which implies a field with linear independent real and imaginary parts.

To get a second and linear independent solution, the field excitation by the beam current has to be considered. There are at least three other approaches (without beam excitation): (1) the power transfer to the beam is replaced by wall losses, (2) the standing wave solution at a different frequency $\omega_2$ is used (with $|\omega_2 - \omega| \ll \omega/Q_L$) and (3) a decaying eigensolution is calculated. Approach (1) needs a (driven) frequency domain solver with surface losses. Approach (2) needs a frequency domain solver without losses (perfect conducting boundary) or a loss-free eigenmode solver with two different cavity geometries (lengths of the FPC waveguide). Approach (3) needs a complex eigenmode solver with waveguide boundaries.

The 3D field map provided by Ref. [31] utilizes the third approach. It describes the cavity fundamental mode including the effects on the fields from both HOM couplers and the FPC. There are different field maps available in Ref. [32], which are calculated for different penetration depths of the coaxial antenna of the FPC. This reflects different values of the loaded quality factor of the cavity.

The field maps are given as a table of sinelike and cosinelike amplitudes, $\vec{E}_b^{\sin}(\vec{r})$ and $\vec{E}_b^{\cos}(\vec{r})$, respectively, for a decaying eigenmode with a backward traveling wave from the cavity into the waveguide.

In general the normalization of eigensolutions is arbitrary in amplitude and phase. For convenience we suppose a phase-normalization such that the electric field energy is maximal at $t = 0$. Therefore the $\vec{E}_b^{\cos}$ part describes the main field that accelerates the particles while the weak $\vec{E}_b^{\sin}$ field is due to the power flow (averaged over one rf period). The origin of the $z$-coordinate can be chosen so that the integrated longitudinal field, observed by a particle with $z = ct$, is maximal. This is approximately the case for an origin in the middle of a cell and in particular for an origin in the middle of a 9-cell cavity.

Under the stated preconditions, the electric field of the backward wave, $\vec{E}_b(\vec{r},t)$, has the following spatiotemporal dependency

$$\vec{E}_b(\vec{r},t) = \vec{E}_b^{\cos}(\vec{r})\cos(\omega t) + \vec{E}_b^{\sin}(\vec{r})\sin(\omega t), \quad (4)$$

with $\omega$ being the angular frequency of the mode. Using Maxwell equations, the electric field of the forward wave $\vec{E}_f(\vec{r},t)$ can be calculated by reversing time as

$$\vec{E}_f(\vec{r},t) = \vec{E}_b^{\cos}(\vec{r})\cos(\omega t) - \vec{E}_b^{\sin}(\vec{r})\sin(\omega t). \quad (5)$$

The overall electric field component for the general case with a given accelerating voltage $V_0$ and phase $\phi$ with respect to the beam can then be calculated as

$$\vec{E}(\vec{r},t) = \Re[(V_0/V_n)e^{i(\omega t+\phi)} \cdot (\vec{E}_b^{\cos}(\vec{r}) - i\mathbf{\Gamma} \cdot \vec{E}_b^{\sin}(\vec{r}))], \quad (6)$$

where $V_n$ normalizes the field to the eigenmode solution of the field map. See Sec. A 1 in the Appendix for a more detailed derivation. For the proposed convenient phase-normalization and origin of the $z$ coordinate, $V_n$ is a real number. The parameter

$$\mathbf{\Gamma} = (A_b e^{i\phi_b} - A_f e^{i\phi_f})/(A_b e^{i\phi_b} + A_f e^{i\phi_f}), \quad (7)$$

describes the ratio between the difference and sum of forward and backward waves at a particular reference plane and corresponds to the negative normalized admittance. The reference plane is chosen to be at a field-node for on-resonance-SW-operation. This is a position that is approximately a multiple of $\lambda/2$ from the tip of the coupler antenna, where $\lambda$ is the rf wavelength. For this choice of reference plane the sum of forward and backward waves [or the denominator of Eq. (7)] is directly proportional to the amplitude of the accelerating field.

Due to the high quality factor of the TESLA cavity wall losses can be neglected. $\mathbf{\Gamma}$ is therefore determined by the amount of beam loading and the detuning of the cavity.

The magnetic field behaves analogously, using similar symmetry properties of the field components, $\vec{B}_f(\vec{r},t) = -\vec{B}_b(\vec{r},-t)$, and it follows that

$$\vec{B}(\vec{r},t) = \Re[V_0/V_n e^{i(\omega t+\phi)} \cdot (\mathbf{\Gamma} \cdot \vec{B}_b^{\cos}(\vec{r}) - i\vec{B}_b^{\sin}(\vec{r}))]. \quad (8)$$

## IV. DISCRETE COUPLER KICKS

A charged particle which traverses the cavity with a given electric and magnetic field configuration on a trajectory with speed $\vec{v}$ at any time $t$ experiences a Lorentz force

$$\vec{F}(\vec{r},t) = q[\vec{E}(\vec{r},t) + \vec{v}(t) \times \vec{B}(\vec{r},t)] = q\vec{V}'(\vec{r},t) \quad (9)$$

where $\vec{V}$ is the effective voltage experienced by the particle. It is plotted in Fig. 6 for an ultrarelativistic particle which traverses the TESLA cavity on axis with the speed of light ($\vec{v}(t) = c\vec{e}_z$) as a function of the longitudinal coordinate $z$. The lower plot indicates the nine-cell geometry of the cavity. The two upper plots point out the influence of the field disturbances caused by the upstream and downstream coupler region, since the particle trajectory is evaluated on axis where the fundamental mode has no transverse components.

The main idea behind discrete coupler kicks (DCK) [12] is to describe the impact of the transverse forces induced by the couplers onto the particle by a discrete kick at the coupler position. The axially symmetrical rf focusing [10] can then be accounted for by using the 2D transfer matrix of the cavity.

The integrated transverse field strength affecting an ultrarelativistic paraxial particle is given by

$$\mathbf{V}_\perp(x,y) = \int dz [\vec{E}_\perp(\vec{r}) + c\vec{e}_z \times \vec{B}(\vec{r})] e^{i\frac{\omega z}{c}} \quad (10)$$

and can be easily separated for the upstream and downstream coupler region for on-axis trajectories by evaluating the integral from/to the center of the cavity to/from infinity. However, if Eq. (10) is evaluated that way for any $x \neq 0$, $y \neq 0$, the offset-dependent edge focusing of the cavity fundamental mode is not compensated and the calculated voltage is not purely induced by the couplers. In order to isolate the integrated transverse fields induced by the couplers

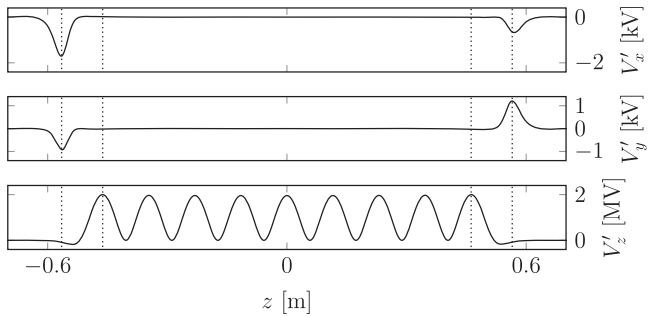

FIG. 6. Horizontal (top), vertical (mid), and longitudinal (bottom) voltage as experienced by an ultrarelativistic charged particle which traverses the cavity on axis.

$$\mathbf{V}_c(x,y) = \mathbf{V}_\perp(x,y) - \mathbf{V}_{RZ}(r) \quad (11)$$

for any $x \neq 0$, $y \neq 0$, the axially symmetrical rf focussing part of the field, $\mathbf{V}_{RZ}$, has to be removed. This can be done by using the 3D field map of the TESLA cavity and extracting the monopole part, for example by averaging the field map over different azimuthal rotations around the cavity axis.

The real part of $\mathbf{V}_c$ corresponds to a net deflection of the bunch centroid, whereas the imaginary part represents a kick which depends on the phase-offset $\Delta\phi = \omega\Delta t$ due to a time offset $\Delta t$ of the individual particle with respect to the reference particle. This time-dependent kick induced by the couplers distorts different longitudinal slices of the beam by a different amount, which results in an increase of the projected emittance [11].

The transverse rf kick is the integrated transverse momentum change relative to the longitudinal momentum of the beam. This kick $\vec{k} = [x', y']$, induced by a coupler, can therefore be calculated as

$$\vec{k}(x,y) = \frac{eV_0}{E_0} \Re\{\tilde{\mathbf{V}}(x,y) \cdot e^{i\phi}\} \quad (12)$$

with $E_0$ being the beam energy at the corresponding coupler region, $V_0$ and $\phi$ the amplitude and phase of the accelerating field, respectively, $e$ the elementary charge and $x$ and $y$ the transverse beam position at the coupler location. $\tilde{\mathbf{V}}$ is the normalized complex kick factor, defined as

$$\tilde{\mathbf{V}}(x,y) = \frac{\mathbf{V}_c(x,y)}{\mathbf{V}_\parallel} \quad (13)$$

with

$$\mathbf{V}_\parallel = \int dz \vec{e}_z \cdot \vec{E}(0,0,z) e^{i\frac{\omega z}{c}} \quad (14)$$

and holds the information of the axially asymmetrical field disturbances induced by the couplers. By taking the field map Eqs. (6), (8) into account, the normalized complex kick factor for the general case of an arbitrary $\mathbf{\Gamma}$ can be written as

$$\tilde{\mathbf{V}}(x,y) = \frac{1+\mathbf{\Gamma}}{2}\tilde{\mathbf{V}}^b(x,y) + \frac{1-\mathbf{\Gamma}}{2}\tilde{\mathbf{V}}^f(x,y) \quad (15)$$

where $\tilde{\mathbf{V}}^b$ and $\tilde{\mathbf{V}}^f$ represent the field disturbances caused by backward and forward traveling wave, respectively. $\tilde{\mathbf{V}}^b$ and $\tilde{\mathbf{V}}^f$ can be calculated directly from a field map, using Eqs. (6), (8) by setting $V_0/V_n = 1$, $\phi = 0$ and $\mathbf{\Gamma} = \pm 1$.

From Eq. (15) directly follows, that $\tilde{\mathbf{V}}$ can be separated in a standing wave part and a reflection dependent part as

$$\tilde{\mathbf{V}}(x,y) = \tilde{\mathbf{V}}^{SW}(x,y) + \mathbf{\Gamma} \cdot \tilde{\mathbf{V}}^{\Gamma}(x,y), \quad (16)$$

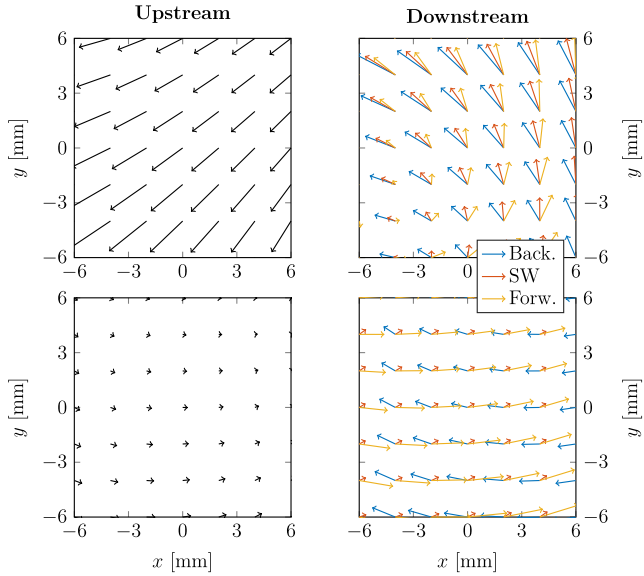

FIG. 7. Real (upper plots) and imaginary part (lower plots) of the normalized complex kick factor for the upstream (left) and downstream coupler region (right) as a function of the transverse coordinates $x$ and $y$. All vectors are scaled by the same amount in order to assure a quantitative comparison. The three colors in the right plots correspond to the cases of pure forward traveling wave (yellow) in the cavity, e.g., no backward wave and $\Gamma = -1$, standing-wave operation (red, $\Gamma = 0$), and pure backward traveling wave (blue, $\Gamma = 1$). Note that the kick induced by HOM couplers does not change for different $\Gamma$. The net effect of the FPC is primarily horizontal.

where $\tilde{\mathbf{V}}^{\text{SW}} = (\tilde{\mathbf{V}}^{\text{b}} + \tilde{\mathbf{V}}^{\text{f}})/2$ and $\tilde{\mathbf{V}}^{\Gamma} = (\tilde{\mathbf{V}}^{\text{b}} - \tilde{\mathbf{V}}^{\text{f}})/2$ hold the field disturbances caused by the sum and the difference of the forward and backward traveling waves, respectively. Only the fields related to the reflection dependent part depend on the parameter $\Gamma$.

The real and imaginary parts of the normalized complex kick factor $\tilde{\mathbf{V}}(x, y)$ as defined in Eq. (16) are plotted in Fig. 7 for both the upstream and the downstream coupler region for different transverse beam positions $x$ and $y$. Different $\Gamma$, thus modes of cavity operation, are evaluated using a field map with an antenna penetration depth of 8 mm.

For the TESLA cavity, the kicks caused by the HOM couplers have the same order of magnitude as that from the power coupler. The kick induced by the upstream HOM coupler does not depend on $\Gamma$, as indicated in Fig. 1. This is due to the fact that the electromagnetic field away from the fundamental power coupler is, to a very good approximation, described by a standing wave and is not affected by the ratio of the forward and backward traveling wave. The $\Gamma$-independent part of the downstream kick relates to the downstream HOM coupler. The $\Gamma$-dependent part relates to the fundamental power coupler, which primarily acts horizontally. Coupler kicks of the FPC for a fixed acceleration gradient have a component that varies with the forward and backward traveling wave, which, for example, would vary as the cavity detuning changes.

## V. NORMALIZED COUPLER KICK COEFFICIENTS

In this section we quantify coupler kicks for different FPC configurations by analyzing the normalized coupler kick coefficients.

It is reasonable to linearize the normalized complex kick factor $\tilde{\mathbf{V}}(x, y)$ around the cavity axis. The zeroth and first order kick $\vec{k}$ on a bunch induced by a coupler as defined in Eq. (12) can therefore be expressed as

$$\vec{k}(x,y) \approx \frac{eV_0}{E_0} \cdot \Re \left\{ \left[ \begin{pmatrix} V_{0x} \\ V_{0y} \end{pmatrix} + \begin{pmatrix} V_{xx} & V_{xy} \\ V_{yx} & V_{yy} \end{pmatrix} \cdot \begin{pmatrix} x \\ y \end{pmatrix} \right] e^{i\phi} \right\}, \tag{17}$$

where $x$ and $y$ are the bunch horizontal and vertical offset at the coupler position. From Maxwell equations it follows that $V_{yy} = -V_{xx}$ and $V_{xy} = V_{yx}$. Thus, coupler kicks are up to first order well described with four normalized coupler kick coefficients $[V_{0x}, V_{0y}, V_{xx}, V_{xy}]$.

It is worth noting that the first order kick acts as a skew quadrupole, and hence can be compensated for constant gradient and $\Gamma$ by adding such a magnet upstream of the cavity, just as the zeroth order dipole kick can be compensated by using a dipole magnet.

As described earlier, the field maps available in Ref. [32] for the TESLA style cavity are calculated for different penetration depths $d = [-12, ..., 12]$ mm of the coaxial antenna of the fundamental power coupler. The corresponding values for the loaded quality factor are $Q_L = [1.81 \times 10^8, ..., 1.36 \times 10^6]$. The normalized coupler kick coefficients $[V_{0x}, V_{0y}, V_{xx}, V_{xy}]$ are calculated for $\tilde{\mathbf{V}}^{\text{SW}}$ and $\tilde{\mathbf{V}}^{\Gamma}$ for the upstream and downstream coupler region with different field maps, thus different values for the loaded quality factor $Q_L$. The coefficients and values for $d$ and $Q_L$ are listed in Tables I, II, and III in the Appendix.

The fields at the upstream coupler are, to a very good approximation, independent of the position of the coupler antenna. However, the impact of the downstream couplers on the transverse beam dynamics is significantly affected by the coupler antenna position.

The $Q_L$-dependence is shown in Fig. 8 for the standing wave and the reflection dependent part in the left and right plots, respectively. The coupler kick is about five orders of magnitude smaller than the longitudinal field. The vertical coefficients $V_{0y}$ and $V_{xy}$ for both waves are insensitive to the antenna position, as it is expected from the geometry. Furthermore they are zero for the traveling wave. However, the horizontal coefficients related to the standing wave depend on the antenna position. For a given $\Gamma$ the overall downstream coupler kick increases with higher $Q_L$.

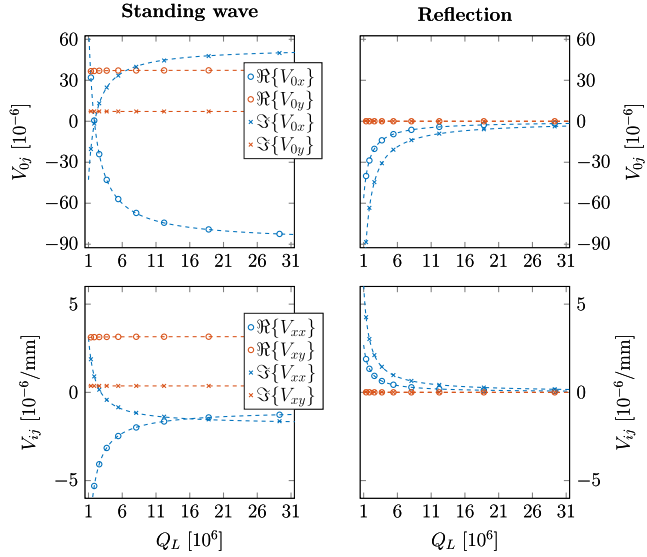

FIG. 8. Normalized complex kick coefficients $[V_{0x}, V_{0y}, V_{xx}, V_{xy}]$ for the downstream coupler region as calculated via Eq. (16) with different field maps, reflecting different loaded quality factors $Q_L$. Plotted are the values related to the standing wave (left) and the reflection dependent part (right). The values are listed in Tables II and III.

The magnitude of the reflection dependent coefficients decreases with higher $Q_L$. This is due to the fact that the transition region in which the traveling wave from the coax changes to a standing wave in the cavity is smaller for larger $Q_L$ [18]. Thus, for very high $Q_L$ values the waves excited on the cavity axis will be standing waves, even in the coupler region.

Real and imaginary parts of most coefficients differ significantly from each other. A variation of the cavity phase will result in coupler kick variations. The imaginary parts of the reflection dependent coefficients exceed the real parts. Therefore, detuning related coupler kick variations are stronger in cavities which are operated off crest, as follows from Eq. (17).

The standard $Q_L$ setting at FLASH is $3 \times 10^6$ whereas it is $4.6 \times 10^6$ at European XFEL. Both $Q_L$ settings are not exactly represented by the field maps. Within the investigated range the $Q_L$-dependence of the coupler kicks is considerable. An appropriate beam dynamics model relying on discrete coupler kicks should therefore use precise coupler kick coefficients. The $Q_L$-dependence of the normalized coupler kick coefficients $V_{ij}$ is well described as

$$V_{ij} = 10^{-6}\left[c_1 + \frac{c_2}{Q_L/10^6 + c_3} + i\left(c_4 + \frac{c_5}{Q_L/10^6 + c_6}\right)\right] \quad (18)$$

with $c_i$ being fit parameters for each coefficient. The dashed lines in Fig. 8 indicate the solutions of Eq. (18) with the fit parameters listed in Tables IV and V in the Appendix.

This coupler kick model allows for a rough estimation of detuning related coupler kick variations. From the rf Eqs. (A7) it follows that for a given accelerating gradient and beam current, the variation $\Delta\Gamma$ and the detuning $\Delta f$ are related according to

$$\Delta\Gamma(\Delta f) = i\frac{2Q_L}{f_0} \cdot \Delta f \quad (19)$$

The horizontal zeroth order coupler kick variation which is related to the reflection dependent part is then given by

$$\Delta k_x^0 = \frac{eV_0}{E_0}\Re\{\Delta\Gamma \cdot V_{0x}^\Gamma \cdot e^{i\phi}\}. \quad (20)$$

Thus, for typical values of $Q_L = 3 \times 10^6$ and $\phi = 0°$ at FLASH it follows that

$$\Delta k_x^0 = \frac{eV_0}{E_0}0.182 \; \mu\text{rad/Hz} \quad (21)$$

which results, for example, in a coupler kick variation of about 3.6 $\mu$rad for a cavity with $V_0 = 20$ MeV, initial beam energy of $E_0 = 100$ MeV and a detuning of $\Delta f = 100$ Hz.

The normalized coupler kick coefficients for the 3.9 GHz cavity are calculated analogously from the field map available at Ref. [33]. The values for orientation 1 (cf. Fig. 5) are listed in Table I in the Appendix.

## VI. IMPLEMENTATION OF COUPLER KICKS IN LINEAR BEAM DYNAMICS MODEL

In this section the implementation of discrete coupler kicks into a linear beam dynamics model is presented, its accuracy for different beam energy is investigated and tracking with ASTRA [34] is used as reference.

Considering linear beam dynamics, the change of transverse coordinates can be written in terms of a matrix formalism as $\vec{u}_1 = R \cdot \vec{u}_0$, with $\vec{u}_1$ and $\vec{u}_0$ representing the particle transverse input and output coordinates $\vec{u}_i = [x, x', y, y']_i$, respectively, and $R$ being the transfer matrix. The transfer matrix of an axially symmetrical rf cavity is given by Ref. [35] and can be found in Eq. (A5) in Sec. A 2 in the Appendix.

The full linear beam transport equation of one cavity equipped with couplers can then be written as

$$\vec{u}_1 = D_0 \cdot \vec{k}_{\text{down}}(D_1 \cdot R_{\text{rf}} \cdot D_1 \cdot \vec{k}_{\text{up}}(D_0 \cdot \vec{u}_0)) \quad (22)$$

where $D_0$ is the drift matrix from the reference positions to the coupler positions, $D_1$ the drift matrix between the couplers and the cavity and $R_{\text{rf}}$ the axially symmetrical cavity transfer matrix. $\vec{k}_{\text{up}}(\vec{u})$ and $\vec{k}_{\text{down}}(\vec{u})$ are the normalized upstream and downstream coupler kick, respectively, at the transverse coordinate $\vec{u} = [x, x', y, y']$, such that $\vec{k}(\vec{u}) = [x, x' + k_x(x, y), y, y' + k_y(x, y)]$. For the drift $D_1$ we use $d_1 = 3.54$ cm, while the cavity length is assumed to be $l_{\rm cav} = 9/2\lambda_{rf} = 1.0377$ m.

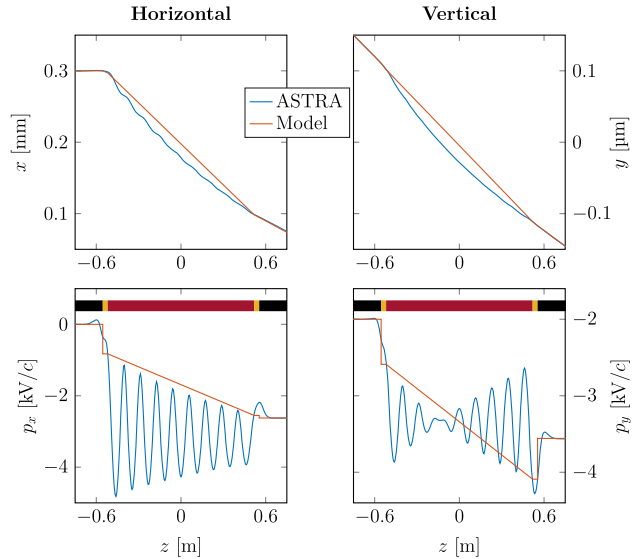

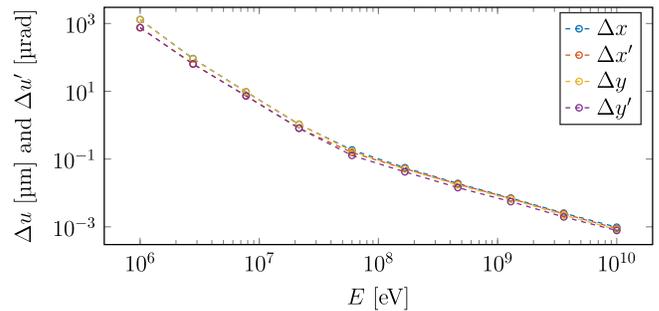

FIG. 10. Comparison between the linear beam dynamics model [cf. Eq. (22)] and ASTRA for different initial beam energies $E$. Plotted are the rms differences $[\Delta x, \Delta x', \Delta y, \Delta y']_{\rm rms}$ as evaluated for $10^4$ particles at different rf amplitudes and phases.

FIG. 9. Particle trajectory through a TESLA cavity as calculated by ASTRA (blue) and the linear beam dynamics model [red, cf. Eq. (22)]. Plotted are the horizontal (left) and vertical (right) positions (upper plots) and momenta (lower plots), respectively, as a function of the longitudinal coordinate $z$. The cavity is centered at $z = 0$ m. The colored bars in the lower plots illustrate the different transfer matrices of the model. Beam initial and final energy is 10 MeV and 24 MeV, respectively.

Figure 9 shows as an example the beam transport through one TESLA cavity as obtained by ASTRA using the 3D field map and by the above described model in blue and red, respectively. Plotted are the horizontal (left) and vertical (right) positions (upper plots) and momenta (lower plots), as a function of the longitudinal coordinate $z$. The cavity is centered at $z = 0$ m and the particle energy is increased from 10 MeV to 24 MeV.

The colored bars in the lower plots of Fig. 9 illustrate the regions at which the different transfer matrices in Eq. (22) are applied. The red bar corresponds to the beam transport region represented by the axially symmetrical rf cavity matrix $R_{\rm rf}$, the yellow bar corresponds to the drift space $D_1$ between the cavity and the couplers and the black line corresponds to the drift $D_0$ between the couplers and the reference positions.

In this example, both the coupler kicks and the rf focusing are sufficiently well described by the linear beam dynamics model. However, a strong dependence of the model accuracy on the beam energy is expected, since the derivation of Eqs. (A5), (17) uses the paraxial approximation, and thus an ultrarelativistic assumption.

The accuracy of the linear beam dynamics model of Eq. (22) for different initial beam energies is calculated as follows. At each energy $E_i$, nine different rf phases between $\phi_{\rm lim} = \pm 20°$ and 7 different rf amplitudes between $V_{\rm lim} = [5 \text{ MV/m}, 25 \text{ MV/m}]$ are evaluated. At each step $[E, \phi, V]_i$, $10^4$ particles with a Gaussian distribution of initial values $[x, p_x, y, p_y]_0$ with $\sigma_{x,y} = 2$ mm and $\sigma_{p_x, p_y} = 5$ keV/c, respectively, are created. For each particle, the output of ASTRA is compared to the corresponding output of Eq. (22).

Results are shown in Fig. 10. Plotted are the rms differences $[\Delta x, \Delta x', \Delta y, \Delta y']_{\rm rms}$, evaluated for all particles and rf amplitudes and phases, respectively, as a function of the initial beam energy $E$.

For beam energies of a few MeV the rms difference for one cavity may reach, for example, several hundred microns. The model is therefore rather unsuitable for accurate beam dynamics calculations regarding the first cavity in a TESLA based injector module as found at FLASH and European XFEL.

Above 100 MeV, however, the rms difference for one cavity is well below 0.1 $\mu$m and 0.1 $\mu$rad, and the beam transport is sufficiently described by Eq. (22). For one module including eight cavities, the rms difference is about 1 mm an for initial beam energy of 5 MeV, where it is about 1 $\mu$m for 100 MeV.

## VII. EXPERIMENTAL VALIDATION OF DISCRETE COUPLER KICK MODEL

In this section the developed beam dynamics model is compared to dedicated experiments at FLASH and European XFEL.

### A. Trajectory response measurements at European XFEL

Trajectory response measurements are a powerful diagnostic tool for linear accelerators [36]. In a linac, the $(i, j)$th element of the trajectory response matrix is defined as the linearized response of a given coordinate ($q_i$) at the $i$th monitor (BPM) to a kick $\Theta_j$ from the $j$th steering magnet [37]

$$\Delta q_i = (R_{i \leftarrow j})_{l,m} \cdot \Delta \Theta_j, \qquad (23)$$

with $R_{i \leftarrow j}$ being the beam transport matrix from point $s_j$ to $s_i$. In the absence of coupling between the two transverse planes, only the elements with indices $l, m = 1, 2$ for the horizontal and $l, m = 3, 4$ for the vertical plane, respectively, are nonzero.

For a given set of rf parameters, the zeroth order coupler kick, that is, $k_x^0 \propto \Re\{V_{0x}\}$, produces a constant kick to the bunch centroid. A trajectory response measurement is consequentially not affected. The first order kick $k_x^1 \propto \Re\{V_{xx} \cdot x + V_{xy} \cdot y\}$, however, depends on the transverse beam position in both planes. At sufficiently low beam energy, it is therefore expected that a trajectory response measurement disagrees considerably with an axially symmetrical beam dynamics model of a cavity and shows coupling between the two transverse planes.

This coupling was observed at the first main accelerating section of the European XFEL, of which a schematic drawing is shown in Fig. 11. After the first bunch compressor (BC0), a linear accelerator (L1) increases the beam energy in four cryomodules, thus 32 cavities, from 150 to 600 MeV. The optics server at European XFEL [38] currently uses ELEGANT [39] for the beam dynamics calculation of $R_{i \leftarrow j}$.

A trajectory response matrix was measured [40] and compared to the calculated response from the optics server. Additionally, the response was calculated with an optics model which includes discrete coupler kicks and calculates the beam transport through each cavity according to Eq. (22). The coupler kick coefficients have been interpolated according to Eq. (18) for $Q_L = 4 \times 10^6$.

Results are shown in Fig. 12. The upper plots are the measured beam trajectories in both transverse planes as excited by an horizontal steering magnet at $s = 86$ m. The lower plots show both the measured trajectory response and the corresponding predictions. The red dots indicate the measurements, whereas the black and blue lines are the calculated response by the default optic server and with the linear model including discrete coupler kicks. The position of the first and last cavity within the string of modules of about 50 m length is indicated by the vertical dotted line in the lower plots.

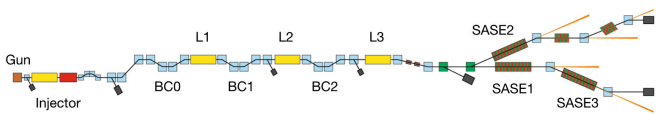

FIG. 11. Schematic layout of the European XFEL (not to scale). The elements shown include 1.3 GHz (yellow) and 3.9 GHz (red) rf sections, undulators (green/red), main dipole magnets (blue), beam distribution systems (green), and beam dumps (black). The main accelerating sections (L1, L2, L3) contain 4, 12 and 84 eight-cavity cryomodules, respectively.

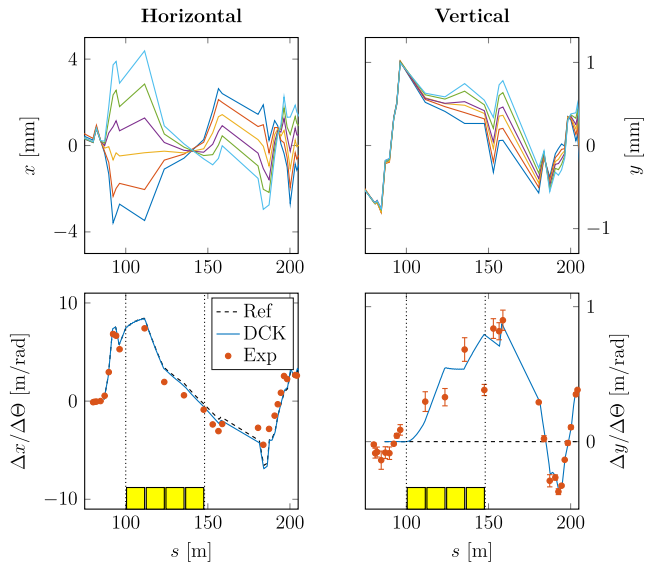

FIG. 12. Trajectory response measurement at European XFEL. The upper plots show the measured horizontal (left) and vertical (right) beam trajectories as excited by an horizontal steering magnet at $s = 86$ m for various magnet strengths. The lower plots show the corresponding trajectory response as measured (red dots), and calculated both by the default optic server (black) and with the linear model including discrete coupler kicks [DCK, blue, cf. Eq. (22)]. The yellow rectangles in the lower row indicate the four L1 cryomodules.

Significant coupling from $x \to y$ occurs in the accelerating modules, which disagrees with the default optics model. A similar optics perturbation including transverse coupling has been also observed at the accelerating sections at FLASH [36,37,41].

The developed model including discrete coupler kicks is able to describe the observed coupling in both planes reasonably well. It can be concluded that the first order coupler kick is described sufficiently by the presented coupler kick model.

### B. Coupler kick variations at FLASH

In this section coupler kick variations related to variations of the forward and backward traveling waves are studied experimentally at FLASH.

The Free-Electron Laser in Hamburg (FLASH) is a high-gain FEL user facility operating in the soft x-ray regime [2,3]. The current layout of FLASH is shown in Fig. 13. The rf setup at FLASH and its effect on the transverse beam dynamics was studied in detail in Ref. [42]. Several cavities are supplied by one high power klystron in pulsed operation with a vector sum rf control. Within the rf flat top, long bunch trains can be accelerated.

An experimental setup in which coupler kick variations within one bunch train can be isolated from the rf focusing at low beam energy was presented in Ref. [44] at the injector module at FLASH. In this section we present a similar experimental setup at high beam energy at the sixth accelerating module.

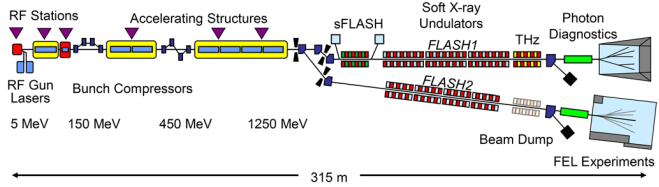

FIG. 13. Schematic layout of the FLASH facility [43].

The cavity field has a limited response to forward power modulations at frequencies much larger than the bandwidth. If the forward rf voltage is modulated at a frequency of $\Omega$, the magnitude of the accelerating field varies as $\Delta V \propto 1/(1 + i(\Omega + \Delta f)\tau)$ with $\tau$ being the decay time of the cavity. For very high modulation frequencies, the response of the accelerating field becomes negligible.

Regarding ultrarelativistic beams, transverse motion due to cavity off-axis fields is insignificant, since transverse magnetic and electric forces compensate each other in axially symmetrical fields. Thus, coupler kick variations are expected to be the dominant source of transverse beam dynamics in this scenario.

Measurements of rf modulated cavity fields and coupler kicks were made using the sixth accelerating module (ACC6) at FLASH. The machine was set up with 400 bunches with a bunch repetition rate of 1 MHz. The rf signals of the forward and backward traveling waves for each cavity were measured. The rf data recorded in the data acquisition system [45] was manually recalibrated according to Ref. [46] in order to remove cross-coupling effects between the forward and backward signals. The beam trajectory was steered to be on axis through the modules and the beam position upstream and downstream ACC6 was measured.

The vector sum of ACC6 was $V_{\text{VS}} = 62$ MV, the bunch charge 0.4 nC and initial beam energy was 600 MeV. The loaded quality factors of the eight cavities were measured as $[3.20, 3.11, 3.17, 3.12, 2.97, 3.02, 3.09, 3.05] \times 10^6$. Each measurement of BPM and rf data was an average over about 100 consecutive bunch trains with 10 Hz repetition rate to deal with short term jitter.

A reference trajectory was measured. Subsequently, the forward power within one bunch train was modulated with different frequencies.

The relative difference of the vector sum of the accelerating gradient $\delta V_{\text{VS}}$ and the parameter $\Delta \Gamma$ computed from the measured forward and backward rf signals with respect to the reference setup are plotted in Fig. 14. The modulation frequencies were 20, 50, 100, and 300 kHz. For high modulation frequencies the variation of the accelerating field becomes small compared to the variation of $\Gamma$ as expected.

In a second step the impact of coupler kick variations on the beam was studied. The measured beam position at the entrance of ACC6, initial beam energy and the rf signals were used to calculate the beam transport through the module. The beam transport for each bunch was evaluated using Eqs. (16), (17), (18), (22) iteratively for each cavity. The measured beam offset at the BPM downstream of the last cavity was compared to the corresponding output of the model function.

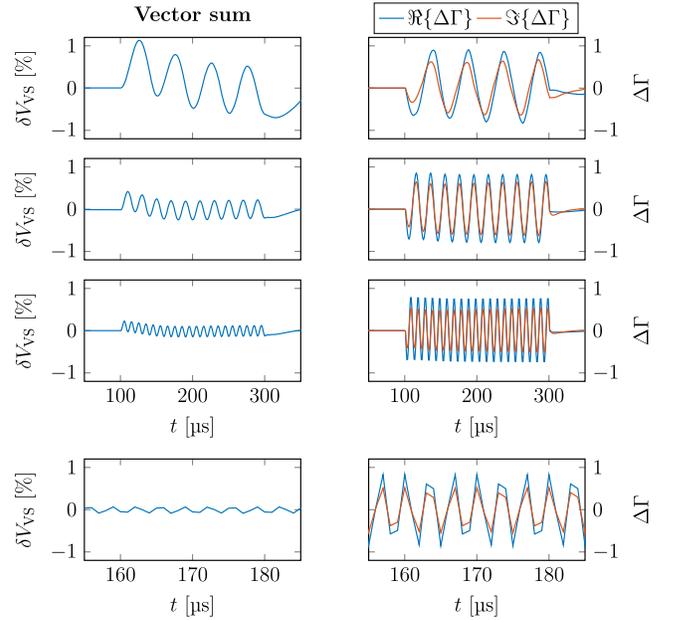

FIG. 14. Variation of the measured vector sum of the accelerating gradient $\delta V_{\text{VS}}$ (left) and the computed parameter $\Delta \Gamma$ (right) at ACC6 at FLASH while applying modulations with different frequencies on the forward power (from top to bottom: 20, 50, 100, and 300 kHz). The rf sampling time is 1 µs.

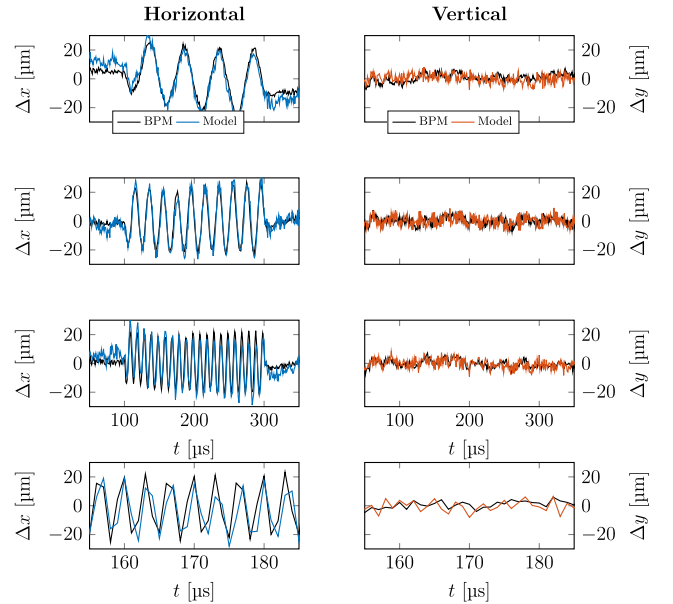

FIG. 15. Experimental observation of intrabunch train coupler kick variations at ACC6 at FLASH. Plotted are the differences between the reference trajectories and the particle trajectories while applying modulations with different frequencies on the forward power (from top to bottom: 20, 50, 100, and 300 kHz, cf. Fig. 14). The bunch spacing is 1 µs. The BPM readout differences (black) and the corresponding model evaluations (colored) are plotted for the horizontal (left) and vertical (right) plane at the exit of the module.

Results are shown in Fig. 15. Plotted are the differences of the horizontal and vertical BPM readouts between the modulated setup and the reference setup and the difference of the equivalent output of the model function. The negligible signal in the vertical plane even for low modulation frequencies points out that the beam dynamics of high energy beams with trajectories close to the cavity axis are dominated by the variation in the coupler kicks.

It can be concluded that the presented coupler kick model is both qualitatively and quantitatively able to reproduce the experimentally generated coupler kick variations.

## VIII. EXPECTED COUPLER KICK VARIATIONS AT EUROPEAN XFEL

In this section we investigate the influence of detuning related coupler kick variations on intrabunch-train trajectory variations at the European XFEL [4–6].

The European XFEL main accelerator consists of 800 TESLA cavities, operated in pulsed mode with an rf flat top of 700 $\mu$s. The long rf pulse structure allows long bunch-trains for the experiments.

The designated pointing stability of the photon beam leads to a stability requirement of 3 $\mu$m maximum trajectory spread within one bunch-train in the undulator section. A conservative estimate predicts worst case beam trajectory perturbations, e.g., from magnet vibrations or spurious dispersion, of about $\pm 100$ $\mu$m assuming a beta function of 30 m [47]. This magnitude of amplitude can be corrected for individual bunches at the entrance to the undulator section by the transverse intrabunch-train feedback system (IBFB) [48]. However, rf-induced trajectory variations have not been considered in the design studies of the IBFB.

The principles of rf induced intrabunch-train trajectory variations have been described in detail in Ref. [42]. At European XFEL, typically 32 cavities with individual operational limits [49] are supplied by one rf power source. The low-level-rf system (LLRF) [20,50] is able to restrict the variation of the vector sum of the amplitude and phase of the 32 cavity fields to below 0.01% and 0.01°, respectively [51]. However, due to the effects of beam loading and Lorentz force detuning, individual cavities have an intrinsic variation of rf parameters within one bunch train. Misaligned cavities in combination with variable rf parameters induce intrabunch-train trajectory variations.

The variation of the amplitude of the accelerating field in each cavity along the bunch train, $\Delta V$, is key in the generation of trajectory variations. For typical machine operation at European XFEL, the amplitude variation is mainly caused by the fact that a common $Q_L$ is used for all cavities yet the gradients are not the same, so the "flattop" voltage along the bunch train is not always flat [52].

LLRF simulations show that the beam loading induced amplitude variation is proportional to the beam current and can reach up to 4 MV/m for the design beam current of 4.5 mA and the actual spread in operational gradients [49] without further $Q_L$-correction.

Beam trajectory variations due to coupler kick variations also have to be considered. High electromagnetic fields in resonators lead to strong Lorentz forces on the walls of these structures. In order to ensure adequate cooling, the thickness and therefore rigidity of the walls cannot be chosen freely. As a consequence, in a pulsed operation mode, the cavities are deformed dynamically in the range of some $\mu$m [20]. This results in a dynamic behavior of the resonance frequency, a Lorentz force detuning (LFD), which scales quadratically with the accelerating field. Due to the high $Q_L$, the LFD within one bunch train is comparable to the bandwidth of the cavity of about 300 Hz [53]. In order to compensate LFD at European XFEL, fine tuning for each cavity will be achieved using double piezoelectric actuators [54]. The residual intrabunch-train detuning of individual cavities is dominated by microphonics and expected to be below 30 Hz.

The European XFEL main linac (L1—L3) increases the electron beam energy from 150 MeV to 17.5 GeV in 100 accelerating modules, thus an energy range in which the ultra-relativistic assumption is reasonable. In order to simulate the effect of trajectory jitter caused by detuning-related coupler kick variations, we use the developed beam dynamics model according to Eq. (22). The coupler kick coefficients for $Q_L = 4.6 \times 10^6$ are listed in Tables I, II, and III. With the rf parameters obtained by Eqs. (A6), the transfer matrices and coupler kick coefficients are calculated for each bunch and each cavity individually. Misalignments are modeled by coordinate transformations. A quadrupole magnet ($k = 0.0642$/m) is located at the downstream end of each module, providing a FODO lattice in the accelerating sections. This is a model from the entrance of L1 to the end of L3 (see Fig. 11) with simplified optics.

As a figure of merit we use the final normalized trajectory variation $\Delta \tilde{u} = \sqrt{\beta_{u,f}(\gamma_{u,z}\Delta u_z^2 + 2\alpha_{u,z}\Delta u_z\Delta u_z' + \beta_{u,z}\Delta u_z'^2)}$, with $\alpha_{u,z}$ and $\beta_{u,z}$ being the Courant-Snyder parameters at position $z$ and $\beta_{u,f}$ the beta function at the observation point, while $u$ stands for the $x$ and $y$ axis. Basically, $\Delta \tilde{u}$ is the rms beam position jitter at a point with a beta function of $\beta_{u,f}$ and zero divergence. For the following analysis we will use $\beta_{x,f} = \beta_{y,f} = 30$ m, which reflects the design lattice at the IBFB downstream the accelerating sections.

For each machine seed the following model parameters were randomly created: variation of the amplitude $\Delta V$ and phase $\Delta \phi$ of the accelerating field (with beam loading) and the detuning $\Delta f$ of individual cavities within the bunch train, and the offset $\Delta u_{\mathrm{cav/mod}}$ and tilt $\Delta u'_{\mathrm{cav/mod}}$ of cavities and modules, respectively. For the range of misalignments we use $\Delta u_{\mathrm{cav/mod}} = \pm 0.5$ mm, $\Delta u'_{\mathrm{cav}} = \pm 0.3$ mrad and $\Delta^{u'}_{\mathrm{mod}} = \pm 0.2$ mrad.

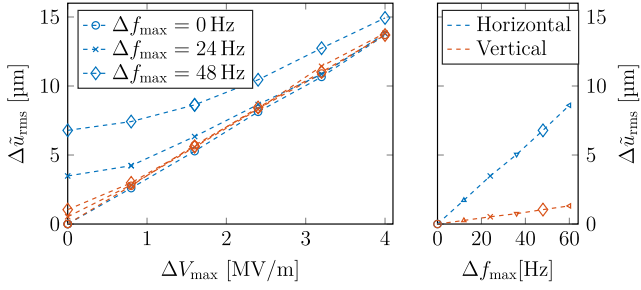

FIG. 16. Normalized rms intrabunch-train trajectory variation $\Delta\tilde{u}_{\mathrm{rms}}$ downstream the European XFEL accelerator as a function of the maximum variation of the amplitude of the accelerating field $\Delta V_{\max}$ (left) for the horizontal (blue) and vertical (red) plane. The different plot marks correspond to different maximum detuning $\Delta f_{\max}$. The right plot shows the trajectory variation for zero amplitude variation as a function of $\Delta f_{\max}$.

The range of the intrabunch-train variation of the rf parameters of individual cavities was increased subsequently. Maximum amplitude and phase variation are considered to be correlated as $\Delta\phi_{\max} = 2\text{m/MV} \cdot \Delta V_{\max}$. For example, a maximum amplitude variation of $\Delta V_{\max} = 4$ MV/m corresponds to a maximum variation of phase $\Delta\phi_{\max} = 8°$. This ratio is found to be typical for regular machine operation. At each parameter step $[\Delta V_{\max}, \Delta f_{\max}]_i$, $10^4$ random machine seeds using a flat distribution were created and the previously described beam transport model was evaluated.

Results are shown in Fig. 16. The left side shows the rms value of the normalized trajectory variation $\Delta\tilde{u}_{\mathrm{rms}}$ at the end of the accelerator as a function of the maximum amplitude variation $\Delta V_{\max}$, evaluated for both planes and three detuning scenarios. The blue and red color correspond to horizontal and vertical plane, respectively, while the different plot marks reflect different detuning limits. The right plot shows $\Delta\tilde{u}_{\mathrm{rms}}$ for zero voltage variation as a function of the maximum detuning $\Delta f_{\max}$.

For small amplitude variations the intrabunch train trajectory variation is dominated by detuning related coupler kick variations and trajectory variations occur mainly in the horizontal plane. For high amplitude variations the trajectory variation is proportional to the amplitude variation, and the influence of detuning related coupler kick variations decreases.

Considering the capability of the IBFB of about $\pm 100$ $\mu$m it can be concluded that coupler kick variations are expected to be a minor concern in multibunch FEL operation at European XFEL. However, for high beam currents, a limitation of beam loading induced amplitude slopes by changing the $Q_L$-setup is advised.

## IX. SUMMARY

Couplers break the axial symmetry of the TESLA cavities and affect the transverse beam dynamics considerably. The concept of first order discrete coupler kicks was presented, including a novel approach to separate the influence of the standing wave from the reflection dependent part. The developed beam dynamics model was compared to standard tracking codes and dedicated experiments. Horizontal-vertical trajectory coupling and detuning related coupler kick variations matched measurements fairly well. Numerical studies regarding European XFEL showed that detuning related coupler kick variations are expected to be a minor concern in regular operation.

## ACKNOWLEDGMENTS

We would like to thank Igor Zagorodnov, Klaus Flötmann and Ángel Ferran Pousa for proofreading and valuable comments. We wish to extend our particular thanks and appreciation to Wolfgang Ackermann, who with great skill and patience calculated the required field maps.

## APPENDIX

### 1. Field calculations

The time harmonic electric field $\vec{E}(\vec{r}, t)$ in a perfect electric conducting cavity that is equipped with couplers can be written as

$$\vec{E}(\vec{r},t) = \Re\{(A_f e^{i\phi_f}\vec{E}_0 + A_b e^{i\phi_b}\vec{E}_0^*) \cdot e^{i\omega t}\}$$

$$= \Re\left\{(A_f e^{i\phi_f} + A_b e^{i\phi_b})e^{i\omega t}\left(\vec{E}_b^{\cos} - i\frac{A_b e^{i\phi_b} - A_f e^{i\phi_f}}{A_b e^{i\phi_b} + A_f e^{i\phi_f}}\vec{E}_b^{\sin}\right)\right\} \quad (A1)$$

with $\vec{E}_0 = \vec{E}_b^* = (\vec{E}_b^{\cos} - i\vec{E}_b^{\sin})^*$. Further

$$\vec{E}(\vec{r},t) = \Re\left[\frac{V_0}{V_n}e^{i(\omega t + \phi)} \cdot (\vec{E}_b^{\cos} - i\boldsymbol{\Gamma} \cdot \vec{E}_b^{\sin})\right] \quad (A2)$$

with $\boldsymbol{\Gamma} = (A_b e^{i\phi_b} - A_f e^{i\phi_f})/(A_b e^{i\phi_b} + A_f e^{i\phi_f})$ and

$$V_n = \int_{-\infty}^{\infty} e^{i\omega(t=z/c)}[\vec{E}_b^{\cos}(0,0,z) - i\mathbf{\Gamma} \cdot \vec{E}_b^{\sin}(0,0,z)]dz \quad (A3)$$

The normalization to $V_n$ is chosen to get

$$V_0 \cos\phi = \int_{-\infty}^{\infty} \vec{E}(0,0,z,t=z/c)dz \quad (A4)$$

A corresponding expression (8) can be derived for the magnetic field. The phase of the eigenmode $\vec{E}_b$ can be chosen so that the electric field energy is maximal for $t = 0$. Then the contribution of $\vec{E}_b^{\sin}$ to the $V_n$ integral is very small, and the integral is practically independent on $\Gamma$. The origin of the $z$ coordinate can be chosen so that $V_n$ is real.

### 2. Cavity transfer matrix

The beam transport in an axial symmetric rf cavity can be written as $\vec{u} = M_{RZ} \cdot \vec{u}_0$ with $\vec{u} = [u, u']$. The matrix elements $R_{ij}$ of the transfer matrix of an axial symmetric cavity are given by Ref. [35] as

$$R_{11} = \cos\alpha - \sqrt{2}\cos\phi\sin\alpha$$
$$R_{12} = \sqrt{8}\frac{\gamma_0}{\gamma'}\cos\phi\sin\alpha$$
$$R_{21} = -\frac{\gamma'}{\gamma}\left(\frac{\cos\phi}{\sqrt{2}} + \frac{1}{\sqrt{8}\cos\phi}\right)\sin\alpha$$
$$R_{22} = \frac{\gamma_0}{\gamma}(\cos\alpha + \sqrt{2}\cos\phi\sin\alpha) \quad (A5)$$

where $\alpha = \frac{1}{\sqrt{8}\cos\phi} \cdot \ln(\frac{\gamma}{\gamma_0})$ and $\gamma_0$ and $\gamma = \gamma_0 + \gamma'$ are the initial and final Lorentz factors, respectively, and $\gamma'$ is the change of $\gamma$ within one cavity. Note that this formula is only accurate for $\gamma \gg 1$.

### 3. LLRF equations

In order to relate the cavity detuning to the ratio of the forward and backward traveling waves, we assume a superconducting cavity operating close to the steady state condition, nearly on crest, with beam loading and a detuning small compared to the resonance frequency. The cavity voltage $\mathbf{V}_0$, the forward- and backward wave $\mathbf{V}_f$ and $\mathbf{V}_b$, and the beam current $\mathbf{I}_b$ are then related as [20]

$$\mathbf{V}_0 = \frac{1 + i\tan\psi}{1 + \tan^2\psi} \cdot [2\mathbf{V}_f + R_L \cdot \mathbf{I}_B]$$
$$\mathbf{V}_0 = \mathbf{V}_f + \mathbf{V}_b, \quad (A6)$$

where the bold letters indicate complex numbers, for example $\mathbf{V}_0 = V_0 \cdot e^{i\phi_0}$.

$\mathbf{I}_B$ is the complex beam current, which is $-2I_{B0}e^{i\phi_B}$ where $I_{B0}$ is the beam current amplitude and $\phi_B$ describes the phase shift of the bunch with respect to a reference phase. The shunt impedance $R_L$ of a cavity without internal losses is determined by the product $Q_{\text{ext}}(R/Q)$ with $Q_{\text{ext}}$ being the external quality factor and $(R/Q)$ being the ratio $V_0^2/(2W\omega_0)$ of the cavity voltage to the stored field energy $W$ and frequency $\omega_0$. The parameter $(R/Q)$ depends only on the shape of a cavity.

The detuning angle $\psi$ is defined as $\tan\psi = 2Q_L\Delta f/f_0$, with $\Delta f$ being the detuning and $f_0$ the resonance frequency of the cavity. Using Eqs. (A6) and assuming a constant beam current, the forward and backward wave can be expressed as a function of the phase difference $\phi_0$ between the bunch and the cavity voltage with an amplitude of $V_0$ and the detuning $\Delta f = f_0 - f$. The parameter $\mathbf{\Gamma}$ follows as

$$\mathbf{\Gamma} = \frac{\mathbf{V}_b - \mathbf{V}_f}{\mathbf{V}_b + \mathbf{V}_f} = \frac{R_L \mathbf{I}_B}{\mathbf{V}_0} + i\frac{2Q_L}{f_0}\Delta f \quad (A7)$$

### 4. Coupler kick coefficients

TABLE I. Coupler kick coefficients of the upstream coupler for the standing wave, $V_u^{\text{SW}}$ (1.3 GHz cavity) and corresponding upstream and downstream $V_u^{\text{SW}}$, $V_d^{\text{SW}}$ and the downstream reflection dependent part, $V_d^{\Gamma}$, for the 3.9 GHz cavity [cf. Eq. (16)].

| Type | $V_{0x}$ [$10^{-6}$] | $V_{0y}$ [$10^{-6}$] | $V_{xx}$ [$10^{-6}$/mm] | $V_{xy}$ [$10^{-6}$/mm] |
|---|---|---|---|---|
| 1.3 GHz $V_u^{\text{SW}}$ | $-57.62 + 9.63i$ | $-41.35 - 0.27i$ | $1.05 - 0.71i$ | $3.37 - 0.02i$ |
| 3.9 GHz $V_u^{\text{SW}}$ | $-94.27 + 249.67i$ | $-30.49 + 143.25i$ | $4.89 - 13.53i$ | $5.63 - 24.26i$ |
| 3.9 GHz $V_d^{\text{SW}}$ | $-515.83 - 34.02i$ | $30.90 + 127.29i$ | $20.72 - 36.41i$ | $5.46 + 23.26i$ |
| 3.9 GHz $V_d^{\Gamma}$ | $-50.12 + 56.31i$ | $3.68 - 1.02i$ | $3.39 - 3.20i$ | $0.62 - 0.29i$ |

TABLE II. Coupler kick coefficients of the downstream coupler for the standing wave, $V_d^{SW}$ [cf. Eq. (16)], as calculated with different field maps of the 1.3 GHz cavity for different values of the loaded quality factor $Q_L$. The values for $Q_L = 3 \times 10^6$ and $Q_L = 4.6 \times 10^6$ reflect the standard $Q_L$-setting for FLASH and European XFEL, respectively, and are evaluated using Eq. (18) and the fit parameters listed in Table IV.

| Coupler position [mm] | $Q_L$ | $V_{0x}$ [$10^{-6}$] | $V_{0y}$ [$10^{-6}$] | $V_{xx}$ [$10^{-6}$/mm] | $V_{xy}$ [$10^{-6}$/mm] |
|---|---|---|---|---|---|
| +12 | $1.36 \times 10^6$ | $32.03 - 20.33i$ | $36.66 + 7.26i$ | $-6.92 + 1.88i$ | $3.12 + 0.37i$ |
| +10 | $1.85 \times 10^6$ | $0.54 - 1.68i$ | $36.84 + 7.23i$ | $-5.31 + 0.90i$ | $3.13 + 0.37i$ |
| +8 | $2.59 \times 10^6$ | $-24.07 + 13.15i$ | $36.97 + 7.20i$ | $-4.07 + 0.15i$ | $3.14 + 0.37i$ |
|  | $3.00 \times 10^6$ | $-31.03 + 17.41i$ | $37.01 + 7.20i$ | $-3.73 - 0.06i$ | $3.14 + 0.37i$ |
| +6 | $3.70 \times 10^6$ | $-42.91 + 24.69i$ | $37.07 + 7.19i$ | $-3.15 - 0.42i$ | $3.14 + 0.37i$ |
|  | $4.60 \times 10^6$ | $-50.30 + 29.29i$ | $37.11 + 7.19i$ | $-2.79 - 0.65i$ | $3.15 + 0.37i$ |
| +4 | $5.41 \times 10^6$ | $-56.95 + 33.43i$ | $37.15 + 7.19i$ | $-2.47 - 0.85i$ | $3.15 + 0.37i$ |
| +2 | $8.07 \times 10^6$ | $-67.10 + 39.86i$ | $37.21 + 7.18i$ | $-1.99 - 1.16i$ | $3.15 + 0.37i$ |
| +0 | $1.22 \times 10^7$ | $-74.28 + 44.49i$ | $37.24 + 7.18i$ | $-1.65 - 1.38i$ | $3.15 + 0.37i$ |
| −2 | $1.88 \times 10^7$ | $-79.18 + 47.73i$ | $37.27 + 7.17i$ | $-1.42 - 1.53i$ | $3.16 + 0.36i$ |
| −4 | $2.93 \times 10^7$ | $-82.47 + 49.96i$ | $37.29 + 7.18i$ | $-1.27 - 1.64i$ | $3.16 + 0.37i$ |
| −6 | $4.59 \times 10^7$ | $-84.63 + 51.50i$ | $37.30 + 7.15i$ | $-1.17 - 1.70i$ | $3.16 + 0.36i$ |
| −8 | $7.23 \times 10^7$ | $-86.04 + 52.52i$ | $37.31 + 7.15i$ | $-1.10 - 1.76i$ | $3.16 + 0.36i$ |
| −10 | $1.14 \times 10^8$ | $-86.91 + 53.24i$ | $37.30 + 7.19i$ | $-1.07 - 1.79i$ | $3.16 + 0.36i$ |
| −12 | $1.81 \times 10^8$ | $-87.54 + 53.62i$ | $37.33 + 7.18i$ | $-1.05 - 1.86i$ | $3.16 + 0.37i$ |

TABLE III. Coupler kick coefficients of the downstream coupler for the reflection dependent part, $V_d^{\Gamma}$ [cf. Eq. (16)], as calculated with different field maps of the 1.3 GHz cavity for different values of the loaded quality factor $Q_L$. The values for $Q_L = 3 \times 10^6$ and $Q_L = 4 \times 10^6$ reflect the standard $Q_L$-setting for FLASH and European XFEL, respectively, and are evaluated using Eq. (18) and the fit parameters listed in Table V.

| Coupler position [mm] | $Q_L$ | $V_{0x}$ [$10^{-6}$] | $V_{0y}$ [$10^{-6}$] | $V_{xx}$ [$10^{-6}$/mm] | $V_{xy}$ [$10^{-6}$/mm] |
|---|---|---|---|---|---|
| +12 | $1.36 \times 10^6$ | $-40.06 - 88.48i$ | $0.02 + 0.37i$ | $1.88 + 4.26i$ | $0.00 + 0.03i$ |
| +10 | $1.85 \times 10^6$ | $-28.71 - 63.36i$ | $0.02 + 0.27i$ | $1.34 + 3.02i$ | $0.00 + 0.02i$ |
| +8 | $2.59 \times 10^6$ | $-20.20 - 44.56i$ | $0.01 + 0.19i$ | $0.93 + 2.11i$ | $0.00 + 0.02i$ |
|  | $3.00 \times 10^6$ | $-17.89 - 39.45i$ | $0.01 + 0.17i$ | $0.83 + 1.87i$ | $0.00 + 0.01i$ |
| +6 | $3.70 \times 10^6$ | $-13.94 - 30.73i$ | $0.01 + 0.13i$ | $0.64 + 1.45i$ | $0.00 + 0.01i$ |
|  | $4.60 \times 10^6$ | $-11.57 - 25.50i$ | $0.01 + 0.11i$ | $0.53 + 1.20i$ | $0.00 + 0.01i$ |
| +4 | $5.41 \times 10^6$ | $-9.44 - 20.79i$ | $0.01 + 0.09i$ | $0.43 + 0.98i$ | $0.00 + 0.01i$ |
| +2 | $8.07 \times 10^6$ | $-6.28 - 13.82i$ | $0.01 + 0.06i$ | $0.29 + 0.65i$ | $0.00 + 0.01i$ |
| +0 | $1.22 \times 10^7$ | $-4.12 - 9.05i$ | $0.01 + 0.04i$ | $0.19 + 0.42i$ | $0.00 + 0.00i$ |
| −2 | $1.88 \times 10^7$ | $-2.67 - 5.85i$ | $0.01 + 0.02i$ | $0.12 + 0.27i$ | $0.00 + 0.00i$ |
| −4 | $2.93 \times 10^7$ | $-1.71 - 3.75i$ | $0.01 + 0.01i$ | $0.08 + 0.18i$ | $0.00 + 0.00i$ |
| −6 | $4.59 \times 10^7$ | $-1.09 - 2.38i$ | $0.01 + 0.01i$ | $0.05 + 0.11i$ | $0.00 + 0.00i$ |
| −8 | $7.23 \times 10^7$ | $-0.69 - 1.51i$ | $0.01 + 0.01i$ | $0.03 + 0.07i$ | $0.00 + 0.00i$ |
| −10 | $1.14 \times 10^8$ | $-0.44 - 0.95i$ | $0.01 + 0.00i$ | $0.02 + 0.04i$ | $0.00 + 0.00i$ |
| −12 | $1.81 \times 10^8$ | $-0.28 - 0.60i$ | $0.01 + 0.00i$ | $0.01 + 0.03i$ | $0.00 + 0.00i$ |

TABLE IV. Fit parameters from Eq. (18) for downstream coupler kick coefficients for the standing wave, $V_d^{SW}$.

| $c_i$ | $V_{0x}$ | $V_{0y}$ [$10^{-6}$] | $V_{xx}$ [$10^{-6}$/mm] | $V_{xy}$ [$10^{-6}$/mm] |
|---|---|---|---|---|
| $c_1$ | −88.292 | 37.322 | −1.001 | 3.160 |
| $c_2$ | 170.294 | −0.921 | −7.878 | −0.074 |
| $c_3$ | 0.058 | 0.030 | −0.027 | 0.510 |
| $c_4$ | 53.925 | 7.170 | −1.830 | 0.364 |
| $c_5$ | −113.160 | 0.079 | 5.335 | 0.010 |
| $c_6$ | 0.170 | −0.464 | 0.085 | 3.004 |

TABLE V. Fit parameters from Eq. (18) for downstream coupler kick coefficients for the reflection dependent part, $V_d^{\Gamma}$.

| $c_i$ | $V_{0x}$ | $V_{0y}$ [$10^{-6}$] | $V_{xx}$ [$10^{-6}$/mm] | $V_{xy}$ [$10^{-6}$/mm] |
|---|---|---|---|---|
| $c_1$ | 0.002 | 0.005 | −0.001 | 0.000 |
| $c_2$ | −49.983 | 0.020 | 2.276 | 0.005 |
| $c_3$ | −0.112 | −0.214 | −0.152 | −0.128 |
| $c_4$ | 0.024 | −0.002 | −0.001 | −0.000 |
| $c_5$ | −110.173 | 0.481 | 5.139 | 0.041 |
| $c_6$ | −0.115 | −0.077 | −0.153 | −0.097 |